\documentclass[draft]{agujournal2019}
\usepackage{url} 
\usepackage{lineno}
\usepackage[inline]{trackchanges} 
\usepackage{soul}

\usepackage{amsmath}
\usepackage{amsfonts}
\usepackage{amssymb}

\draftfalse

\journalname{Geophysical Research Letters}

\DeclareMathOperator\arctanh{arctanh}
\DeclareMathOperator\erf{erf}

\begin{document}

\title{Osmosis drives  explosions and methane release in Siberian permafrost}

\authors{Ana M. O. Morgado\affil{1}, Luis A. M. Rocha\affil{1}, Julyan H. E. Cartwright\affil{2,3}, and Silvana S. S. Cardoso\affil{1}}

\affiliation{1}{Department of Chemical Engineering and Biotechnology, University of Cambridge, Cambridge CB3 0AS, UK}

\affiliation{2}{Instituto Andaluz de Ciencias de la Tierra, CSIC--Universidad de Granada, 18100 Granada,  Spain}

\affiliation{3}{Instituto Carlos I de F\'{\i}sica Te\'orica y Computacional, Universidad de Granada, 18071 Granada,  Spain}

\correspondingauthor{Ana M. O. Morgado}{amom2@cam.ac.uk}
\correspondingauthor{ Julyan H. E. Cartwright}{julyan.cartwright@csic.es}

\begin{keypoints}
\item Surface ice-melt water can migrate downward driven by the osmotic pressure associated with a cryopeg, a lens of salty water below.  
\item  Overpressure can cause the frozen soil to crack resulting in mechanical explosion. 
\end{keypoints}

\begin{abstract}
Mysterious craters, with anomalously high concentrations of methane, have formed in the Yamal and Taymyr peninsulas of Siberia since 2014.  While thawing permafrost owing to climate warming promotes methane releases, it is unknown how such release might be associated with explosion and crater formation. A significant volume of surface ice-melt water can migrate downward driven by osmotic pressure associated with a cryopeg, a lens of salty water below.  Overpressure reached at depth may lead to the cracking of the soil and subsequent decomposition of methane hydrates, with implications for the climate.
\end{abstract}

\section*{Plain Language Summary}
We show how osmosis drives explosions and methane release in Siberian permafrost. We anticipate that as well as being of direct relevance to permafrost researchers, this work will be of interest to a large number of people involved in climate change research, because the mechanism we uncover of osmotic pumping leading to permafrost explosions has potentially grave consequences involving the release of methane presently locked up in hydrates. 

\section{Introduction}\label{sec1}

\begin{figure}
	\centering{\includegraphics[width=0.8\textwidth]{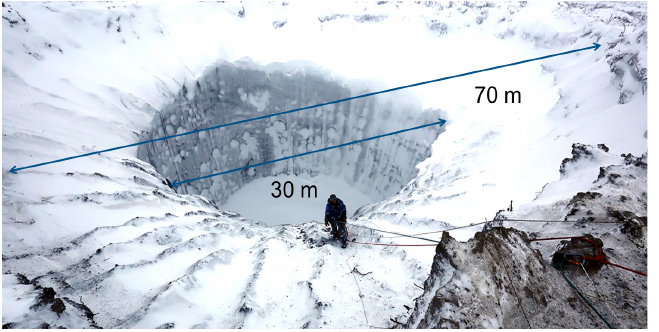}}
	\centering{\includegraphics[width=0.8\textwidth]{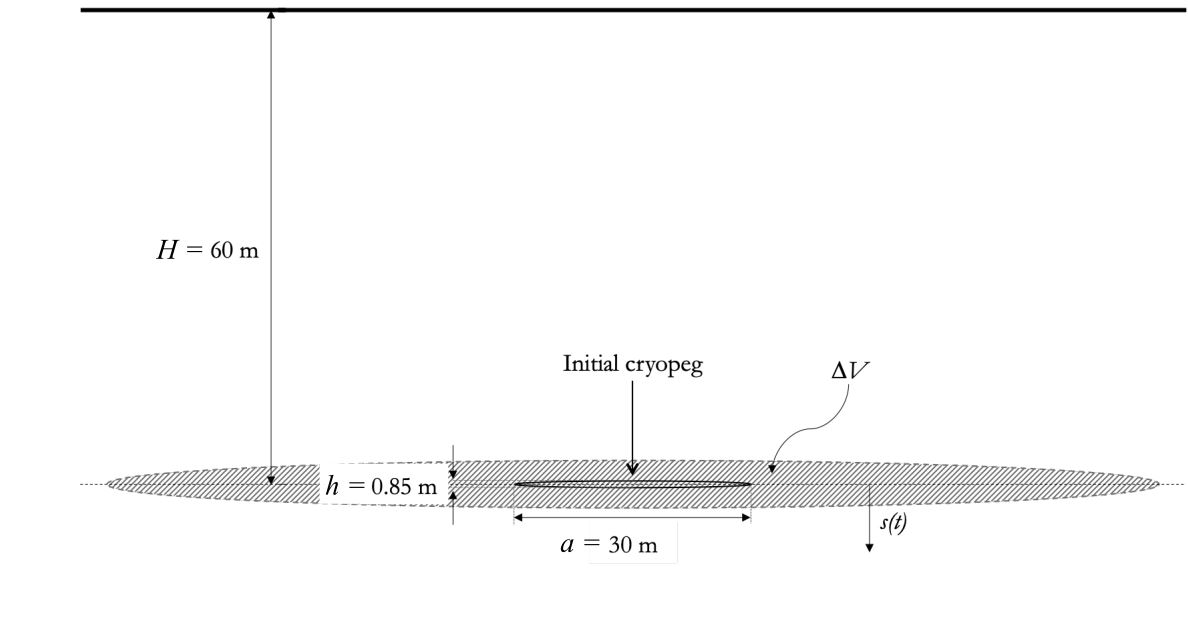}}
	\centering{\includegraphics[width=0.8\textwidth]{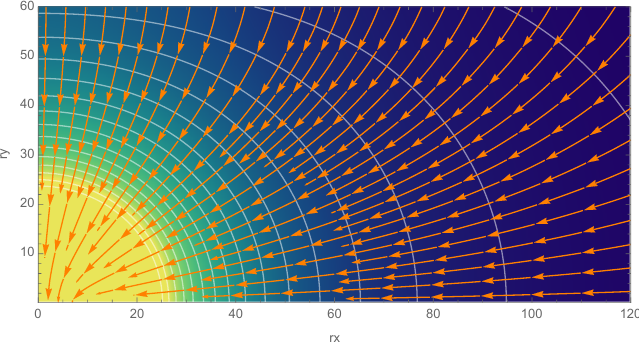}}
	\caption{(Top) The first observed crater in the permafrost, detected in 2014 in the Yamal Peninsula, Siberia. Source: National Geographic.
		(Centre) Schematic representation of proposed cryopeg inflation.
		(Below) 2D Streamlines illustrating flow of water from the surface into the cryopeg (The $y$-axis is an axis of symmetry).}
	\label{fig:crater}\label{fig:graph}
\end{figure}

In 2014, a mysterious new crater was detected in the Yamal Peninsula, Siberia; Figure~\ref{fig:crater} \cite{Leibman2014}. The form of the crater plus the ejecta surrounding it suggest that it was produced by an explosive process.
Since then, scientists and locals have located several other such features \cite{dvornikov2019}.  Anomalously high concentrations of methane measured during expeditions to the Yamal crater \cite{Leibman2014}  suggest that this hydrocarbon is being released to the atmosphere as a result of its formation.
Methane is a potent greenhouse gas and its presence in the atmosphere thus contributes to climate change.  A substantial amount of methane is trapped deep in the permafrost in the form of methane hydrates, ice-like solids that are crystallized mixtures of water and methane. The methane presumably originates from the biogenic conversion of organic matter \cite{kraev2017,Froitzheim2021} and some estimates predict that methane in hydrates may be the largest source of hydrocarbons on Earth \cite{buffett2000}. Hydrates are stabilized by high pressure and low temperature, and they are sensitive to changes in the environment \cite{ruppel2017}. Rising Arctic permafrost temperatures have been recorded in recent years \cite{Kurylyk2014,Romanovsky2010} and it is known that for subaerial permafrosts surface temperature is the major factor affecting the ground thermal regime \cite{Williams1989}. The timescale of this dissociation process of hydrates has frequently been considered to be millennia \cite{turetsky2019}, but convective fluid flow driven both by osmosis and buoyancy can accelerate this timescale to decades \cite{Cardoso2016}.

The Yamal Peninsula is located in a region of  permafrost with a thickness of 180--300~m \cite{Leibman2014, Olenchenko2015}. The soil is predominantly composed of clay loams with a maximum porosity of 35\% \cite{Leibman2014, Yakushev2000}. The ground has high ice content of up to 85\% of the pore volume \cite{Yang2019}; liquid water fills the remaining pore volume \cite{Leibman2014, Olenchenko2015, Chuvilin2000,Yakushev2000}. The existence of unfrozen water as thin films adsorbed onto the grain surfaces in frozen, clay-water environments has been demonstrated in the laboratory  \cite{Anderson1967,Hoekstra1965,Burt1976}. A high concentration of mineral ions in solution, expelled by growing ice crystals, ensures that the film surrounding the clays remains unfrozen. In frozen silicate--water systems, these unfrozen, interfacial water films have thicknesses of about 5--70~$\mu$m \cite{Anderson1967}. The mobility of these water films was suggested by  Hoekstra \cite{Hoekstra1965} and demonstrated by Burt \& Williams \cite{Burt1976}. Anderson \cite{Anderson1967} noted the occurrence of electrical and thermal osmosis within frozen earth materials at temperatures as low as -10 \textdegree C as the most relevant evidence of the fluidity and continuity of the water films. Frozen-soil permeabilities in Siberia of the order of $10^{-15}~$m$^2$ have been measured \cite{Boike} (see also \cite{kane1983,kane2001}), suggesting much larger water films.  Clays of the type found in the Yamal Peninsula are known to exhibit osmotic effects for film thicknesses of a few microns \cite{Bresler1973}.  This combination of a relatively permeable clay soil and osmotic effects can drive a significant subsurface flow of water \cite{neuzil2000,Neuzil2009}.

The existence of cryopegs, lenses of high salinity water, has been reported at depths below 50~m in the  Yamal Peninsula  \cite{Yakushev2000,Streletskaya1998}, i.e., just above the metastable hydrate layer. These cryopegs developed after the Pleistocene sea regression due to saline exclusion during sediment freezing, and are brines, composed mainly of sodium chloride of concentration of 7 to 150 g L$^{-1}$, typical thicknesses of 0.5--12 m, and lateral extensions up to 300~m  \cite{Streletskaya1998}. At the  surface, the permafrost is overlain by an active layer, which thaws during summer, around June to August, and subsequently refreezes, owing to the seasonal variations in air temperature \cite{Williams1989}.  The presence of the highly mineralized unfrozen water in the cryopegs and this seasonal thawing of the active layer suggest that osmotic pressure may promote water flow from the surface to depth.

The absence of reports of emission of light or combustion products \cite{Leibman2014} during the explosions leading to the crater formation in Yamal suggests it is  unlikely that these craters are the product of a chemical reaction; a physical mechanism is required to explain them.
 In this work, we show that the thawing in the active surface layer associated with increasing Arctic temperatures, coupled with water migration to depth driven by osmosis into a cryopeg, can lead to cracking of the permafrost with concomitant destabilization of the methane hydrates. We demonstrate below that, although the temperature increase of the soil at the depth of the hydrate layer is small, the predicted pressure rise from the accumulation of water can trigger soil fracture within a timescale of  decades.  Soil fracturing in turn drastically reduces the pressure at depth, which can thus trigger hydrate decomposition leading to the release of methane gas in a violent physical explosion.  

Global temperatures are now rising, which may increase the frequency of these explosive events. The temperature in the Arctic has increased twice as fast as the rest of the globe, causing shorter periods of ice coverage and longer periods over which the explosion-induced transport of methane to the surface can occur \cite{Shakhova2015}. Owing to the greenhouse effect methane has on the atmosphere, this creates a positive feedback loop to climate change \cite{turetsky2019}.

This work attempts to explain the mechanism behind the formation of these craters through a physical model based on the osmotic flow towards a cryopeg present at depth.

\section{Evolution of pressure and volume of water in a cryopeg}

Consider  water flow in the porous permafrost soil, extending from the surface to a cryopeg.  The flow in the small channels between the ice and the solid grains is driven by the osmotic pressure established by the  concentration of brine in the cryopeg. From Darcy's law \cite{Turcotte2014} we expect the superficial speed
\begin{equation}
	u=-\frac{k}{\mu} \nabla {p},
	\label{darcy} 
\end{equation}
where $p=P-p_a-\rho_m g (H - z)$ is the reduced pressure,  
$\rho$ denotes density and subscript $m$ refers to the porous medium saturated with ice. 
Incompressibility requires further that $\nabla \cdot u=0$.
The cryopeg is idealized as an oblate spheroid with width $a$ and thickness $h$ at a depth  $H$; Figure~\ref{fig:graph}. It has an initial volume $V_0$ and porosity $\phi_0$. It initially has salinity $C_{s0}$, which imposes an osmotic pressure according to van't Hoff's equation $p_c = 2 \sigma R T  C_{s0}$ \cite{van1901osmotic}, driving water into the cryopeg at a flow rate $Q$. This increases the volume of the cryopeg $V= 4/3 \pi a^2 h$, and the pressure $p$ within it. Eventually, the pressure may reach a critical value causing the structure to rupture.
The porosity is 
\begin{equation}
	\phi  = \frac{V_w}{V_s + V_w} = \frac{V_w}{V}.
	\label{eq:ellipsoid}
\end{equation}

The pressure profile is mathematically equivalent to the electric potential around a charged object of the same geometry \cite{landau}. Thus, for an ellipsoidal cryopeg, the approximate axisymmetric pressure distribution is
\begin{equation}
	p = \frac{1}{\beta}\ln\left( \frac{V_w}{V_{w0}} \phi_0 + \left(1 - \phi_0\right) \right).
	\label{eq:pressure}
\end{equation}
Here $p$ is the reduced pressure and $p_{c}=2 \sigma RT c_s$ the osmotic pressure on the cryopeg boundary; $\sigma$  is the osmotic reflection coefficient \cite{Bresler1973,Neuzil2009}. The pressure and volume of the cryopeg $V$
are constrained by the geo-mechanical properties of the surrounding soil, $\beta$.  For an oblate spheroid  \cite{Amoruso2009},
${\mathrm{d} V}/{\mathrm{d} p_c} =\beta V$,
with 

\begin{equation}
    \beta = \frac{1}{2 \mu_m} \frac{1-2 \nu_m}{1+\nu_m} \left[ \frac{a}{h} \frac{4(1-\nu_m^2)}{\pi (1-\nu_m)}-3 \right],
    \label{eq:beta}
\end{equation}

where
$\nu_m$ and $\mu_m = E/\left(2(1 + \upsilon)\right)$ are respectively the Poisson ratio and the shear modulus of the ice and soil mixture. 

Coupling this mechanical constraint with the conservation of volume of water and salt in the cryopeg,  $\mathrm{d}V_w/{\mathrm{d}t} = Q$ and $c_s={c_{s0}V_0}/{V}$, gives the rate of increase of volume of water as
\begin{equation}
	Q = \frac{dV_w}{dt} = Q_s \left( \frac{V_{w0}}{V_w} - \gamma p \right),
	\label{eq:volume}
\end{equation}
where
\begin{equation}
	Q_s = \frac{k}{\mu}2\pi A a\left( 1 - \frac{2 \eta_0}{3 \left( 2H/a \right)^3} \right)
	\label{eq:flow}
\end{equation}
with
\begin{equation}
    A = \frac{p_c}{-\cot^{-1}\left( \sinh \eta_0 \right) + \cot^{-1}\left( 2H/a \right)};
    \label{eq:parameterA}
\end{equation}
here $\eta_0 = \arctanh (h/a)$ and $\gamma = 1/p_c$.

The driving pressure difference $p_{c} - p$ is mainly determined by the local salt concentration in the cryopeg, particularly for early and intermediate times, which we expect to be uniform owing to natural convection driven by the inflow of fresh water at the bottom. We have therefore assumed that the driving pressure is constant over the cryopeg boundary. The magnitude of the osmotic effect in a frozen clay soil has not been studied.  For loam soils and low salinity of 0.08--0.10N, an osmotic coefficient of up to 0.16 has been reported \cite{letey1969}.  While a higher salinity in Yamal \cite{streletskaya2017} would decrease this value \cite{Neuzil2009}, we expect ice with its impermeable behaviour to increase the osmotic coefficient.  

\begin{table}[tb]\small
	\caption{Base case parameters used.}
	\label{table:base_case}
	\centering\begin{tabular}{lllll}
		\hline
		permafrost permeability & $k$ &  $5 \times 10^{-15}$ & m\textsuperscript{2} & \cite{Boike} \\
		viscosity of water & $\mu$ & $1.8 \times 10^{-3}$ & Pa.s & \\
		ideal gas constant & $R$ & 8.314 & J K\textsuperscript{-1} mol\textsuperscript{-1} & \\
		permafrost temperature & $T$ & 267.3 & K &  \cite{Williams1989} \\
		permafrost Poisson ratio & $\nu$ & 0.33 & - & \cite{schulson1999} \\
		permafrost Young's modulus & $E$ & $9 \times 10^{6}$ & Pa & \cite{schulson1999} \\
		cryopeg initial width & $a$ & 15 & m &  \cite{Streletskaya1998} \\
		cryopeg initial thickness & $h$ &  1.05 & m  &  \cite{Streletskaya1998} \\
		cryopeg depth & $H $ & 60 & m & \cite{ Yakushev2000,Streletskaya1998} \\
		permafrost initial porosity & $\phi_0$ &  0.3 & -  & \cite{Kurylyk2014} \\
		permafrost osmotic coefficient & $\sigma$ & 0.16 & -  &  \cite{letey1969}\\
		cryopeg initial  NaCl salinity  & $C_{s0}$ & 10 & g L\textsuperscript{-1} & \cite{Streletskaya1998} 
		\\
		permafrost cracking pressure & & 0.17 & MPa  & \cite{Williams1989} \\
		active layer thawing period & & 3 &  months/year & \cite{Williams1989} \\
		\hline
	\end{tabular} \\
\end{table}

\section{Heat flow, pressure change and methane hydrate dissociation}

The timescales for osmotic water flow and convective heat flow are, respectively,
$\tau_{water}= {H \phi}/{u_0}$
and
$\tau_{heat}= {H}/{u_0}$.
Assuming $C_{s0} = 10$ g L\textsuperscript{-1} and applying equation \ref{darcy}, the osmotic speed at $t=0$ is estimated to be $u_0=1.8 \times 10^{-7}$ ms$^{-1}$, so that $\tau_{water} = 0.33$ yr and $\tau_{heat}=10.6$ yr, indicating that the water flow is much faster than the heat flow. The effect of the latter on the thermodynamic state at the top of the hydrate layer depends on the relative evolution of pressure and temperature in the cryopeg,  
which can be estimated from the rate of temperature change at the surface \cite{Leibman2014}, $\Gamma=0.2 $ K yr$^{-1}$, and the pressure change to give
$dT/dP=0.03$ K MPa$^{-1}$. This indicates that the temperature increase in the cryopeg is very slow compared to that of pressure, so we expect the thermodynamic state at the top of hydrate layer to consist of subcooled ice and hydrates for a permeability of $\kappa=5 \times 10^{-15}$ m$^2$, typical of the Yamal soil. The expected small temperature change lends support to our simple model  considering only the pressure field.

\section{Results}\label{results}

\begin{figure}[tbp]
	(i) \\ \includegraphics[width=\linewidth]{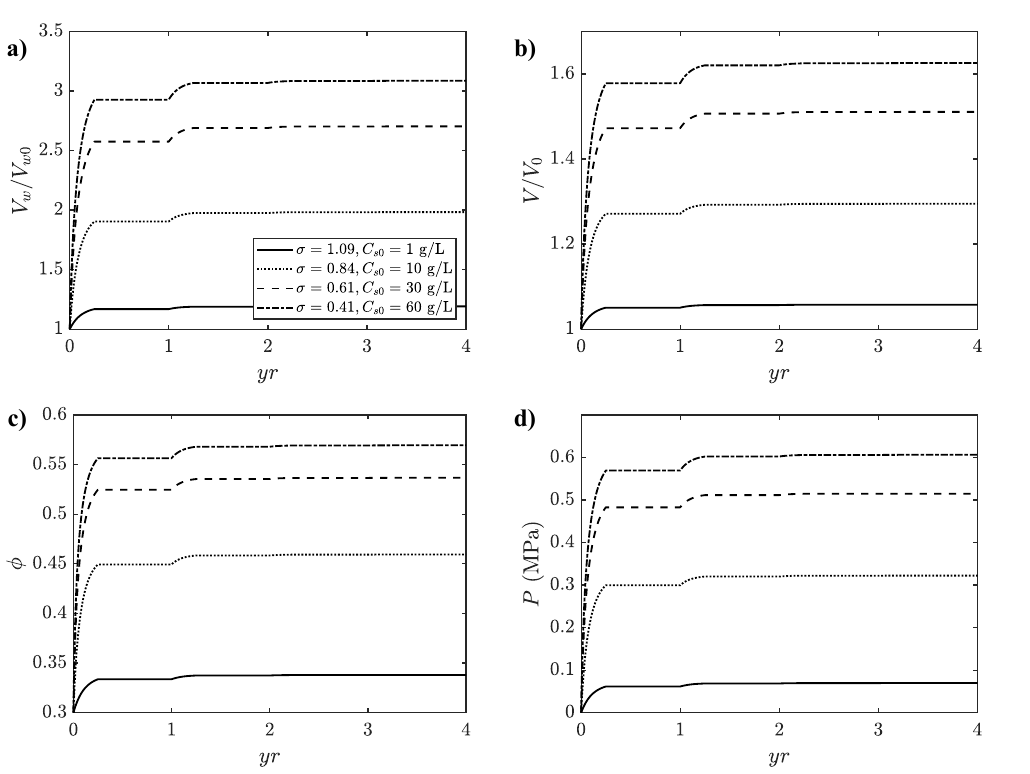}
	(ii) \\\includegraphics[width=\linewidth]{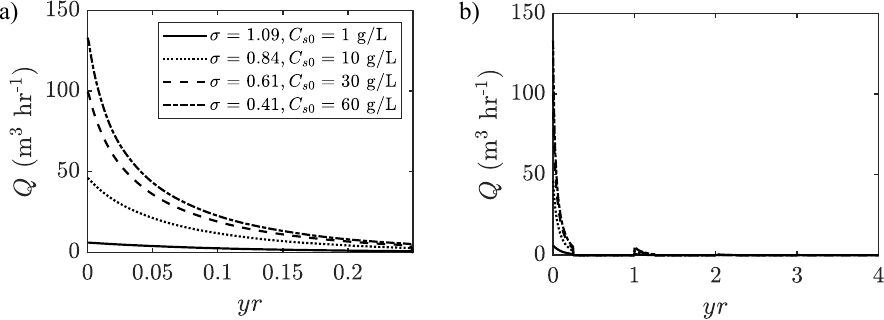}
	\caption{(i) Evolution of the cryopeg under the base case conditions given in Table~\ref{table:base_case} plus different levels of cryopeg salinity.
	(ii) Volumetric flow rate $Q$ against time for the baseline conditions,  a) zooming in on one summer and
 b) showing a 4 year period.
	}
	\label{fig:base}\label{fig:flow_rate}
\end{figure}

\begin{figure}[tbp]
	\includegraphics[width=\textwidth]{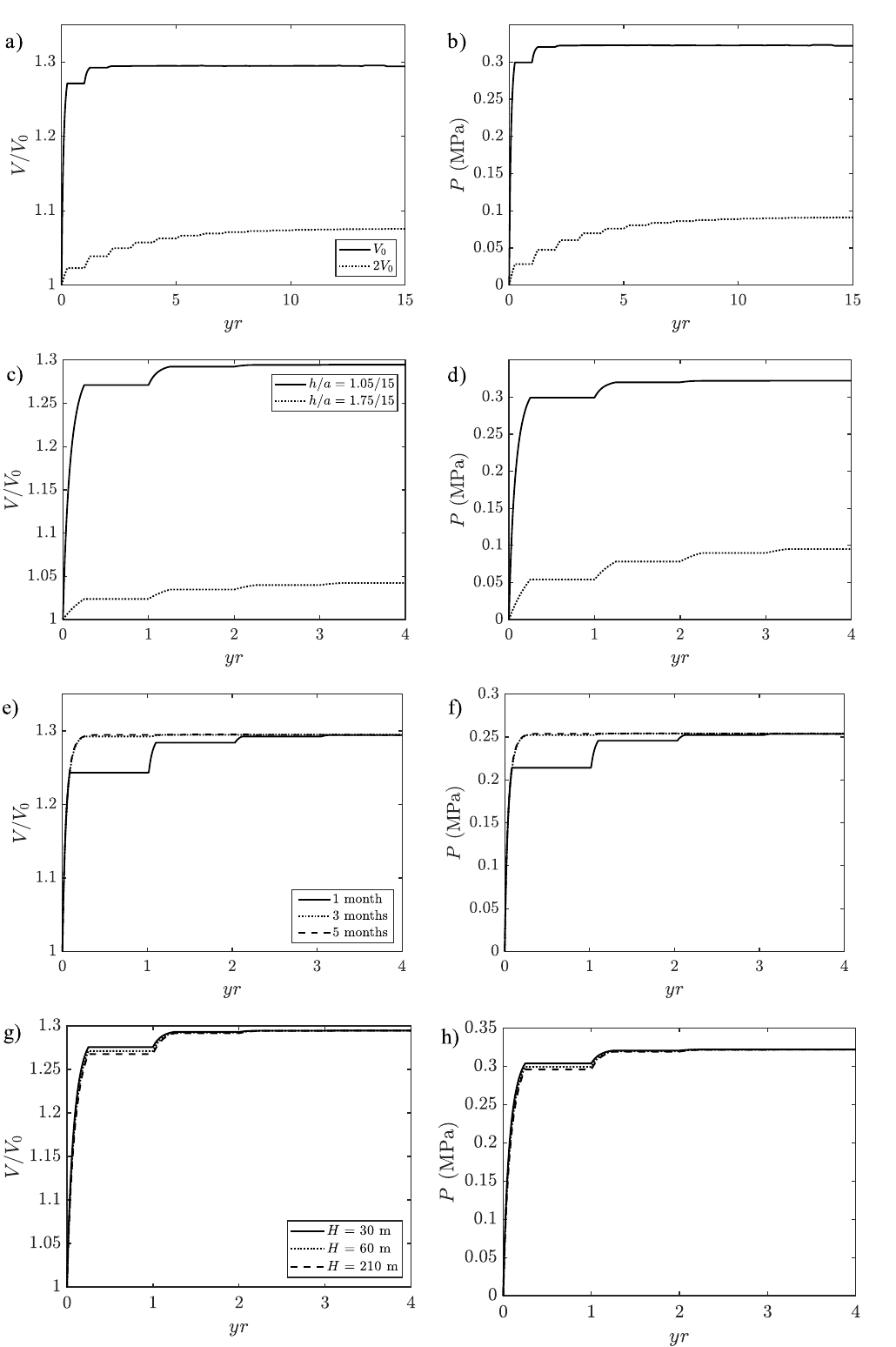}
	\caption{Cryopeg volume and pressure changes for  
		(a,b) differently sized cryopegs;
		(c,d)  different cryopeg aspect ratios;
		(e,f) different yearly thawing periods;
		(g,h) different cryopeg depths.}
	\label{fig:volume_pressure}
\end{figure}

\begin{figure}[tbp]
	(i) \\\includegraphics[width=\linewidth]{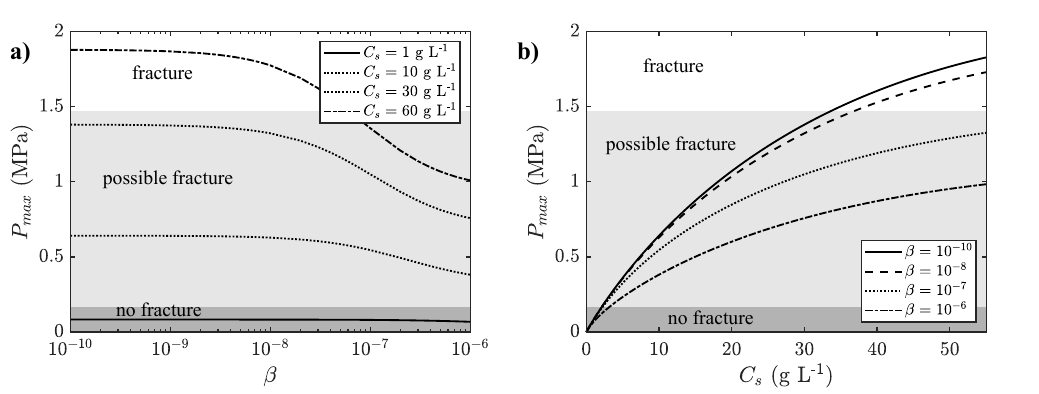}
	(ii) \\	\centering{\includegraphics[width=\textwidth]{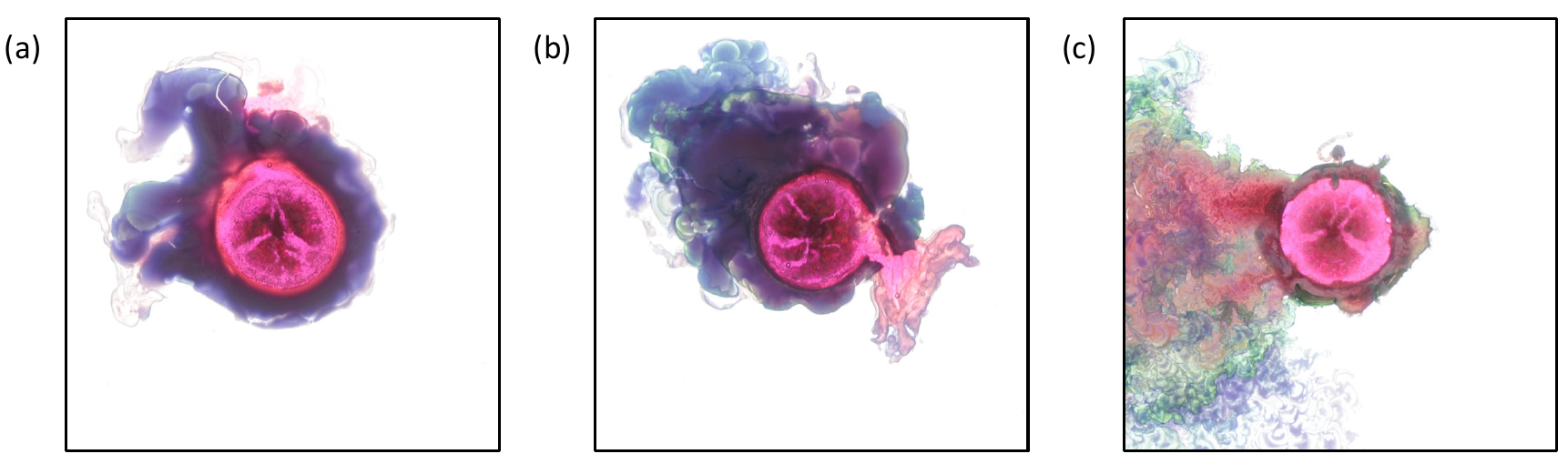}}
	\caption{(i) Maximum pressure in a cryopeg as a function of 
		(a) $\beta$;
		(b) salinity.
		(ii) Analogue laboratory experiments demonstrate osmosis and explosion in a Hele-Shaw cell. The colors seen correspond to coordination states of the colbalt ion involved.
	}
	\label{fig:regimes}\label{growth}
\end{figure}

In permafrost that is not covered by ice, ground surface temperature undergoes fluctuations owing to seasonal variations of air temperature. As a consequence, permafrost is overlaid by an active layer, which undergoes seasonal thawing in summer and subsequent refreezing \cite{Makogon1997}.  At Yamal this layer currently has an average thickness of 1~m \cite{Leibman2014}, but it has been extending downwards in the last decades as a consequence of warmer air temperatures and longer summers \cite{Vasiliev2020}.  
The evolution of the volume of water and overpressure in the Yamal cryopeg, obtained from the numerical integration of equation  \eqref{eq:volume}, is shown in Fig.~\ref{fig:base} for the base case parameters listed in Table~\ref{table:base_case}  with different levels of cryopeg salinity.  It is seen  that, under these conditions,  within 4 years the volume of water accumulated in the cryopeg tends to up to 3 times the original volume, causing an overpressure of magnitude up to 0.6~MPa.  
In Fig.~\ref{fig:flow_rate} we give the volumetric flow rate $Q$ against time for the baseline conditions. 
We are  considering the dynamics to be active during the summer months, three months per year here.  Thus the curves present bursts and plateaux since the evolution of the cryopeg system is not continuous. 
In Fig.~\ref{fig:volume_pressure} we illustrate the volume and pressure changes for further variations of this base case: differently sized cryopegs; different cryopeg aspect ratios; different yearly thawing periods, and different cyropeg depths.

A question  to be answered is whether the maximum pressure for any range of parameters encountered in nature is greater than the fracture strength of the soil. 
Results are presented  in Fig.~\ref{fig:regimes} for the maximum pressure from \eqref{eq:pressure} as a function of salinity and $\beta$.  
Figure~\ref{fig:regimes}a shows the maximum pressure in the cryopeg versus $\beta$ for different concentrations. We took $b = 5$ when estimating $\sigma$, which leads to  high values for the osmotic efficiency. In order to do this we fitted data from Bresler  \cite{Bresler1973} to the expression $y = k_1(1 - \erf(k_2 b \sqrt{N}))$,
where $N$ is the salinity normality, $k_1 = 1.210$, $k_2 = 0.0941$ and $b = 5$. 
Fig~\ref{fig:regimes}a indicates that fracture is possible with $C_s$ 3 g/l and low $\beta$, as well as with $C_s$ 10 g/l and all the range of $\beta$ explored. This needs a compressed layer of sediment or very low permeability, included in the assumption $b=5$.
In Fig.~\ref{fig:regimes}b we have plotted $P_{max}$ versus $C_s$ for different $\beta$. We considered concentrations up to 60 g/L. For $b = 5$, $\sigma$ decreases more slowly than before, and as a result the osmotic pressure  keeps increasing with increasing $C_s$.
The maximum overpressure reached for all $\beta$ between 10$^{-6}$ and 10$^{-10}$ and for salinities 3 and 10 g L$^{-1}$
is within the interval of the possible fracture of ice-rich frozen silt soils, 0.17--1.3 MPa \cite{Williams1989}, and for $\beta\leq 10^{-8}$ it exceeds the upper limit of their fracture strength,  suggesting that osmotic accumulation of water within the cryopeg is capable of producing  soil fracture.  
In most permafrost settings heterogeneity of the ice content and therefore the permeability of the medium can have an important effect on osmotic convection. As the pressure in the surrounding soil increases, through water accumulation in the cryopeg, the local permeability may decrease. Such an effect would decrease the flow of water and therefore extend the period over which the cryopeg inflates. The estimate of the time period to explosion for the basic homogeneous system presented here constitutes
therefore a lower bound for  heterogeneous frozen soils.

Kizyakov et al.\ \cite{Kizyakov2015} reported a maximum diameter of 45--58 m, and a maximum height of 5--6 m for a pingo ---  a large mound of ice, which develops in permafrost due to segregation of massive ice lenses \cite{Williams1989} --- noted prior to the explosion that formed the Yamal crater \cite{Olenchenko2015}. Taking this deformation as half of an axisymmetric ellipsoid, a volume of 7.6$\times 10^3$ m$^3$ of displaced material at the surface is obtained,   larger than the cryopeg volume increase from the modelling above.  This  suggests the accumulation of a large volume of methane gas near to the surface after cracking of the soil but prior to its explosion at the surface. The cracking mechanism is then followed by the physical explosion as methane hydrate depressurizes and decomposes into methane gas.
A similar mechanism in volcanism involving a bubble of trapped gas  as a means to 
rupture a sealed melt reservoir and initiate a magmatic eruption has been termed advective overpressuring \cite{pyle1995,woods1997}, and the formation and rupture of a bubble --- although of carbon dioxide rather than methane --- has been proposed as a cryovolcanism mechanism involved in the formation of the Yamal crater \cite{buldovicz2018}.

\section{Laboratory experiments}\label{experiments}

To demonstrate this osmotic pumping  and explosion mechanism,  it is instructive to consider   analogous  laboratory experiments using a chemical garden reaction \cite{barge2015}.  These reactions form semipermeable precipitate membranes at the interface between reactants such as silicate and metal salts. The interactions between osmosis and buoyancy can lead to self assembling patterns. A Hele-Shaw cell, consisting of two parallel plates separated by a thin gap, may be used as an analogue for two-dimensional flow in the porous soil.  In Fig.~\ref{growth}, a pellet of compressed cobalt chloride powder was placed in the cell and an aqueous solution of sodium silicate was injected into the environment surrounding it.  As the pellet dissolves, the cobalt and silicate ions react according to  $ \rm{Co}^{2+}\rm{(aq)}  +  \rm{SiO}_3^{2-} \rm{(aq)} \rightarrow \rm{CoSiO}_3 \rm{(s)} $, to form a semi-permeable membrane of cobalt chloride.  The membrane separates  two aqueous solutions: the saturated solution of cobalt chloride inside and the dilute silicate solution outside, which establishes a steep concentration difference across the precipitate membrane. As a result, an osmotic pressure develops, which drives water flow inwards, across the membrane, and thereby increases the internal pressure.  During the early stages of the formation process, the precipitation membrane has low mechanical strength and eventually cracks under the growing internal pressure.  At later times the pressure increases owing to the continuing osmotic pumping, and the membrane 
undergoes a stronger fracture process, with the pellet shattering and explosive ejection of CoCl$_2$ solution \cite{ding2019}.  
While this membrane is formed after a chemical reaction, the evolution of the structure is dependent on the physical mechanism of osmotic flow. This is thus an example of osmosis leading to an explosion after some time.

Figure~\ref{growth} illustrates the formation of the membrane and its subsequent cracking for a 1.5 M Na$_2$SiO$_3$ solution.
In this analogue laboratory experiment for the interaction of osmotic pumping \cite{ding2020} and cryopeg inflation,
the buoyant release of depleted salt solution can be associated with early fracturing and partial release of the cryopeg fluid, while the later stronger cracking at larger pressure mimics  rapid decompression when the cracks open to the atmosphere.

\section{Discussion}\label{discussion}

Despite the mediatic buzz caused by the discovery of several mysterious craters in the Siberian peninsulas of Yamal and Taymyr, in 2014, a physical model for the process at their origin is yet to be proposed. In this study, we set out to investigate whether the leading explanation offered for the process---that the increasing warmer Artic temperatures and consequent thawing of the permafrost promoted methane releases that led to a violent explosion and crater formation---was physically plausible.  
In porous permafrost soil, ice-melt water at the surface can flow in small channels between the ice and the solid grains, driven by the osmotic gradient established between the surface water and the brine in the cryopeg. Modelling the cryopeg as an oblate spheroid, a mathematical expression for the volume of water in the cryopeg \eqref{eq:volume} is obtained. 

Maximum pressure in a cryopeg is plotted as a function of salinity and factor $\beta$, to understand whether,for any range of geological parameters, it would surpass the soil fracture strength. The results show that soil fracture may occur both under $C_s= 3$ g L$^{-1}$ and low $\beta$ conditions, and $C_s=10$ g L$^{-1}$ and the entire range of $\beta$ investigated, requiring a compressed sediments layer with very low permeability, compatible with the assumption of $b = 5$ initially made. 
Further analysis of the maximum pressure as function of salinity (as high as 60 g L$^{-1}$), for different values of $\beta$, showed that for $b = 5$, porosity decreases at a slower rate, and thus the osmotic pressure increases with increasing salinity concentration. The maximum pressure obtained for all $\beta$ values between 10$^{-6}$ and 10$^{-10}$, and salinities in the 3--10 g L$^{-1}$ range, falls within the interval of possible fracture of ice-rich frozen soils. For $\beta \leq 10^{-8}$ it exceeds the upper limit of their fracture strength, suggesting that osmotic accumulation of water within the cryopeg can lead to soil fracture. 

To illustrate this osmotic pumping and explosion mechanism, i.e., the interaction of osmotic pumping and cryopeg inflation, an analogous laboratory experiment using a chemical garden reaction was undertaken.  Two periods of release are observed: first, during the early stages of the formation process, when the membrane has low mechanical strength, eventually cracking due to the growing internal pressure, and then a stronger fracture process, owing to the continuing osmotic pumping, and consequent pressure increase. The initial release can be associated with early fracturing and partial release of the cryopeg fluid, while the later stronger cracking at larger pressure mimics rapid decompression when the cracks open to the atmosphere. 
The results of this work demonstrate that overpressure reached at depth, owing to the migration and accumulation of a significant volume of surface ice-melt water in the cryopeg, can exceed the soil fracture strength and cause the frozen soil above to crack. If such fractures propagate to the surface, a rapid reduction of pressure at depth can result in methane hydrate decomposition, and consequent methane gas release, and mechanical explosion.

The Polar regions are not filled with these craters. We may infer that the explosion is a rare phenomenon. Ice sheets covered the regions in which explosions have occurred until very recently in geological terms. The Yamal peninsula was glaciated in the Early Weichselian, c.\ 90--80ka BP,  if perhaps the limit was north of it later \cite{svendsen2004}.  Although cryopegs have been present during all this period, they have remained stable. The mechanism we have described above functions over a much shorter time frame to increase cryopeg pressure, yet 
generally the situation has been stable, and  only in some exceptional cases do we get these explosions. The trigger may be climate warming and the increase in depth of the active layer that provides for the amount of liquid water needed to blow up a cryopeg.
The annual rate of active layer thawing reported in 2006--2013 \cite{Leibman2014} 
is sufficient to support the estimated osmotic flow of water migrating towards the cryopeg during the thawing season,  allowing for evaporation of water at the surface in addition to transport to depth.

\section{Conclusions}\label{conclusions}

The  mechanism for groundwater flow and pressure increase proposed here can therefore cause soil fracture and relief of pressure at depth, which in turn may enhance the release of methane gas from hydrate decomposition, and ultimately lead to violent explosion.  The transport of methane hydrates from depth to the surface during explosion can significantly accelerate their decomposition and promote the direct release of methane gas into the atmosphere. The projected formation of these features and the concomitant methane emissions from gas hydrate decomposition should be taken into consideration in predictions of potential feedbacks leading to the warming of Earth's climate.

A similar mechanism  may lead to destabilization of the methane-hydrate deposits within the seabed \cite{Archer2009}. Pockmarks on the ocean floor are thought in some instances to be the result of explosive release of mainly methane gas \cite{andreassen2017}. These crater-like features range from 50 to over 500~m diameter and are found in fine-grained sediments such as silts and clays \cite{Hovland2007}.  They have been noted in lake beds on the Yamal peninsula \cite{yakushev2018}.
It is thought that  post-glacial rapid climate warming and  associated degradation of sub-seabed permafrost ice lenses might be at the origin of a period of explosive activity during which many pockmarks formed \cite{Hovland2007}.  Current climate warming could enhance an osmotic pumping mechanism within the seabed, as proposed above. A growing activity of explosions may therefore ensue, with associated release of methane into the ocean waters and atmosphere.

Further work may explore the typical volume of gas that is released in these explosions, and their potential height into the atmosphere. Furthermore, it is relevant to assess the number of currently existing cryopegs.
The model may also be verified by checking if the explosions occur most often during or just after summer.

\section*{Open Research Section}

The mathematical model developed in this work is fully presented in the text.

\acknowledgments

A.M.O.M. gratefully acknowledges funding from the Vice-Chancellor's and Newnham College Scholarship, awarded by the Cambridge Commonwealth, European and International Trust. 
L.A.M.R. gratefully acknowledges funding from the Funda\c{c}\~{a}o para a Ci\^{e}ncia e Tecnologia (FCT), Portugal (grant SFRH/BD/130401/2017).




\end{document}